\documentclass[iop]{emulateapj}
\usepackage{multirow}
\begin{document}

\title{Testing Galactic Magnetic Field Models using Near-Infrared Polarimetry}
\author{Michael D. Pavel, D. P. Clemens, \& A. F. Pinnick}
\affil{Institute for Astrophysical Research}
\affil{Boston University, 725 Commonwealth Ave, Boston, MA 02215}
\email{pavelmi@bu.edu; clemens@bu.edu; apinnick@bu.edu}

\slugcomment{Accepted for Publication in The Astrophysical Journal}
\shorttitle{Testing Galactic Magnetic Fields}
\shortauthors{Pavel, M. D., Clemens, D. P., \& Pinnick, A. F.}

\begin{abstract}
This work combines new observations of NIR starlight linear polarimetry with previously simulated observations in order to constrain dynamo models of the Galactic magnetic field. Polarimetric observations were obtained with the Mimir instrument on the Perkins Telescope in Flagstaff, AZ, along a line of constant Galactic longitude ($\ell=150\degr$) with 17 pointings of the $10\arcmin \times 10\arcmin$ field of view between $-75\degr < b < 10\degr$, with more frequent pointings towards the Galactic midplane. A total of 10,962 stars were photometrically measured and 1,116 had usable polarizations. The observed distribution of polarization position angles with Galactic latitude and the cumulative distribution function of the measured polarizations are compared to predicted values. While the predictions lack the effects of turbulence and are therefore idealized, this comparison allows significant rejection of A0-type magnetic field models. S0 and disk-even halo-odd magnetic field geometries are also rejected by the observations, but at lower significance. New predictions of spiral-type, axisymmetric magnetic fields, when combined with these new NIR observations, constrain the Galactic magnetic field spiral pitch angle to $-6\degr\pm2\degr$.

\end{abstract}

\keywords{ISM: dust - ISM: magnetic fields - Polarization - Radiative transfer}

\section{Introduction}
The Galactic magnetic field is an important aspect of the interstellar medium, but even its large-scale structure is still uncertain. What symmetry properties does the field exhibit? Is the magnetic field axisymmetric, or are there field direction reversals at some radii? How is the magnetic field sustained? As a first step to addressing these questions, this work compares new near-infrared (NIR) starlight polarimetry to predictions of the polarization properties for different magnetic field geometries \citep{P11}. Through strong observational constraints, the mechanism that sustains the Galactic magnetic field can also be constrained. In addition to the large-scale symmetry of the Galactic magnetic field, these observations are used to constrain the spiral-type magnetic pitch angle of the Galaxy.

Magnetic fields remain one of the most difficult aspects of the nearby universe to probe. Every method for measuring the interstellar magnetic field requires some special circumstance (e.g., a background polarized source for Faraday rotation, high column density for polarized thermal emission or Zeeman splitting, a background star for interstellar polarization of background starlight). To date, the study of the Galactic magnetic field has been dominated by radio wavelength Faraday rotation studies. However, by combining many different techniques, observational constraints can be placed on the structure of the Galactic magnetic field which will drive, for example, dynamo theories forward. 

Over the last few decades, progress has been made in understanding the Galactic magnetic field, through, especially, improvements in numerical simulations \citep{E92,B92,B93,FS2000,K02,K03,M08,HWK09,M10}. However, while our understanding of fundamental dynamo physics grew, there were few observational predictions. Observational constraints of large-scale magnetic field models have almost exclusively relied on Faraday rotation of emission from Galactic pulsars and polarized extragalactic sources \citep{S68, Man74, RK89, MFH08} which only probe the line-of-sight component of the magnetic field. \citet{H96} used optical polarimetry to constrain the Galactic magnetic pitch angle (defined as $p=tan^{-1}(B_r / B_\phi)$ for a spiral-like magnetic field pattern), but, beyond this, starlight polarization has not been quantitatively employed as a tool for constraining the large-scale Galactic magnetic field structure.

Polarization of background starlight has been an astrophysical tool since the pioneering works of \citet{HI49} and \citet{HA49}, and early theoretical work linked this optical polarization with interstellar magnetic fields \citep{DG51}. As currently understood \citep{L07}, interstellar polarization arises from dichroic extinction of unpolarized starlight passing through regions of elongated dust grains which are directionally aligned with their long axes preferentially perpendicular to the local magnetic field. Assuming background stars emit all polarizations equally (i.e., unpolarized), photons polarized parallel to a grain's long axis see the largest grain cross section and are preferentially extincted. The result is linear polarization whose stronger electric field direction is parallel to the direction of the magnetic field, as projected on the sky.

Much of the previous starlight polarization work has been done at optical wavelengths where the polarization signal is strongest \citep{SMF75}. However, optical wavelengths suffer appreciable dust extinction, especially through the Galactic disk, and typically cannot probe stars farther than 2-3 kpc \citep{F02,H08}. At longer wavelengths, starlight is less extincted by dust and polarizations can be measured to stars far beyond optical limits.

In addition to extinction, nearby magnetic structures can dominate the observed optically-traced morphology. Loop I is a supernovae remnant of $116\degr $ diameter on the sky, approximately 130 pc from the Sun, centered at $\ell=329\degr$, $b=17.5\degr$ \citep{B73}, and prominently seen in the optical polarization compilation of \citet{H2000}. This nearby object obscures the signature from the large-scale magnetic field structure. Faraday rotation studies of the Galactic magnetic field may also be affected by similar foreground objects \citep[e.g., supernova remnants:][]{W2010}. By observing in a direction without any obvious disturbances, the quiescent Galactic magnetic field may be reliably probed by NIR starlight polarimetry to distances of several kiloparsecs \citep{C12a}.

\citet{P11} used existing dynamo-driven Galactic magnetic field models and empirical dust distributions to predict the NIR observational signatures of different magnetic field models: S0 (even), A0 (odd), disk-even halo-odd (DEHO), and simple analytic axisymmetric. From these predictions, the shape of the curve of polarization Galactic position angle (hereafter GPA\footnote{Throughout this work, polarization position angles will be measured in the Galactic coordinate system, measured East ($+\ell$) from Galactic North (+$b$).}) with Galactic latitude ($b$) and cumulative distribution functions (CDFs) of the normalized degree of polarization ($P$) were proposed as tools for testing predictions of the large-scale structure of the Galactic magnetic field against observations.

Here, the predictions of \citet{P11} are tested against observations of NIR starlight polarimetry with the goal of constraining possible large-scale magnetic field geometries. In \S 2, the observations and data reduction are described. A summary of the simulated polarization measurements from existing Galactic dynamo and dust models is presented in \S 3, along with a comparison between the predicted and observed polarimetric properties. The results of this analysis are discussed in \S 4, and conclusions are presented in \S 5.

\section{Observations and Data Reduction}
Observations were made with the Mimir instrument \citep{C07} in H-band ($1.6\mu$m) linear imaging polarimetry mode on the 1.8m Perkins Telescope outside Flagstaff, AZ, on several nights from 2007 through 2009. Mimir uses a cold, stepping half-wave plate (HWP) and a cold, fixed wire grid to analyze starlight polarization across a $10\arcmin \times 10\arcmin$ field of view. The detector is a $1024 \times 1024$ InSb Aladdin III array.

The observations were taken along a line of constant Galactic longitude, $\ell=150\degr$, for $-75\degr<b<10\degr$ in steps of $\Delta b = 5\degr$ to enable comparison with the \citet{P11} predictions. Observations taken at $b=-5\degr$ were not used because the bright star HD 23049 contaminated the images, instead the neighboring $b=-6\degr$ field was substituted. Galactic latitudes $b=0\degr,\, -35\degr,\, -50\degr,\,$ and $-55\degr$ were not observed because of poor weather and the Perkins telescope is not able to observe North of $b=10\degr$ or South of $b=-75\degr$ at this Galactic longitude due to telescope mount limits. Supplemental measurements were obtained near the Galactic midplane at $b=\pm1.25\degr$ and $\pm2.5\degr$ to provide higher latitude resolution there.

\begin{deluxetable}{ccccc}[t]
	\tabletypesize{\footnotesize}
	\tablecaption{\label{obs-table}Observations and Starlight Polarization Detections}
	\tablewidth{0pt}
	\tablecolumns{5}
	\tablehead{\colhead{Galactic} &
						 \colhead{UT Date} &
						 \colhead{Significant} &
						 \colhead{Marginal} &
						 \colhead{Upper} \\
						 \colhead{Latitude} &
						 \colhead{Observed} &
						 \colhead{Detections\tablenotemark{a}} &
						 \colhead{Detections\tablenotemark{b}} &
						 \colhead{Limits\tablenotemark{c}}\\
						 \colhead{(1)} &
						 \colhead{(2)} &
						 \colhead{(3)} &
						 \colhead{(4)} &
						 \colhead{(5)} \\
						 }
	\startdata
		$-75 \degr $ & 2009 Nov 27 & 0 & 7 & 4\\
		$-65 \degr $ & 2009 Dec 1 & 1 & 5 & 1 \\
		$-60 \degr $ & 2009 Nov 27 & 0 & 14 & 6 \\
		$-45 \degr $ & 2009 Nov 27 & 1 & 14 & 5 \\
		$-40 \degr $ & 2009 Dec 1 & 1 & 10 & 7 \\
		$-30 \degr $ & 2009 Dec 1 & 4 & 13 & 6 \\
		$-25 \degr $ & 2009 Nov 27 & 2 & 12 & 7 \\
		$-20 \degr $ & 2007 Nov 24 & 4 & 29 & 12 \\
		$-15 \degr $ & 2007 Nov 24 & 6 & 46 & 28 \\
		$-10 \degr $ & 2007 Nov 24 & 8 & 57 & 24 \\
		$-6 \degr $ & 2009 Dec 1 & 34 & 54 & 14 \\
		$-2.5 \degr $ & 2007 Nov 25 & 76 & 108 & 8 \\
		$-1.25 \degr $ & 2007 Nov 25 & 60 & 117 & 12 \\
		$1.25 \degr $ & 2008 March 23 & 37 & 97 & 23 \\
		$2.5 \degr $ & 2007 Nov 25 & 51 & 93 & 16 \\
		$5 \degr $ & 2008 March 23 & 19 & 52 & 5 \\
		$10 \degr $ & 2007 Nov 25 & 28 & 56 & 16 \\
		\hline
		& Total & 332 & 784 & 194
	\enddata
	\tablenotetext{a} {Significant detections defined as $P/\sigma_P > 3$}
	\tablenotetext{b} {Marginal detections defined as $ 1.25 \leq P/\sigma_P < 3$.}
	\tablenotetext{c} {Upper limits defined as $P/\sigma_P < 1.25$ and and $\sigma_{P} < 1.0\%$.}
\end{deluxetable}

For each pointing, the HWP is rotated to 16 different position angles, equivalent to five measurements at an instrument position angle (IPA) of $0\degr$ and four IPA measurements each at $45\degr, 90\degr,$ and $135\degr$. This is done toward six different sky dither positions, following a rotated hexagon pattern on the sky, for a total of 102 images per pointing. Each target field consisted of four sets of observations with 10s integrations per HWP position, for a total of 68 minutes of integration time per field. Integration times were set to ensure that the polarimetric uncertainty at $H \leq 14$ was less than 1\%. The dates of each Galactic latitude observation are shown in column (2) of Table \ref{obs-table}. Most of the stars found have complementary NIR photometry from 2MASS \citep{S06}.

Polarization standards were also observed for calibration. For the 2007 observations, the unpolarized star HD 42807 \citep{G74} and polarized standard star HD 30675 \citep{W92} were observed for calibration. For all other observations, the Galactic globular cluster NGC 5466 was observed as a collection of unpolarized stars and the polarization standard stars Elias 14, 22, and 25 in the Scorpius field \citep{W92,C12b} were observed. Detector linearity was characterized by observing the constant surface brightness of an in-dome screen for a range of exposure times, and each of the 16 HWP position angles had a unique dome flat measured, described in detail in \citet{C12a}.

Data reduction was performed with two custom IDL software packages (available through the Mimir website\footnote{http://people.bu.edu/clemens/mimir/software.html}), whose details can be found in \citet{C12a}. The first software package performs quality testing and applies linearity, dark, and polarimetric flat field corrections. The second package calculates astrometric solutions for each image, coadds images, extracts photometry, calculates H-band starlight polarizations, and applies instrumental polarization corrections. Since each of the fields was observed with four pointings at six dither positions per pointing in sixteen unique HWP positions per dither, a total of 24 ten-second exposures were taken through each HWP position of each field (the HWP IPA$=0\degr$ position had 48-ten second exposures for each field). For each field, these 24 exposures were coadded to form sixteen master HWP images per field. PSF-assisted aperture photometry \citep{C12a} of stars in these master HWP images were used to calculate the polarization properties of each star. The final output is a catalog of stellar polarizations (polarization percentage, equatorial position angle, Stokes U, Stokes Q, and associated uncertainties) with Mimir H-band photometry and 2MASS JHK$_{s}$ photometry, where available. The sixteen master HWP images were also coadded to create deep photometric images for each field.

\begin{figure}[ht]
	\centering
		\plotone{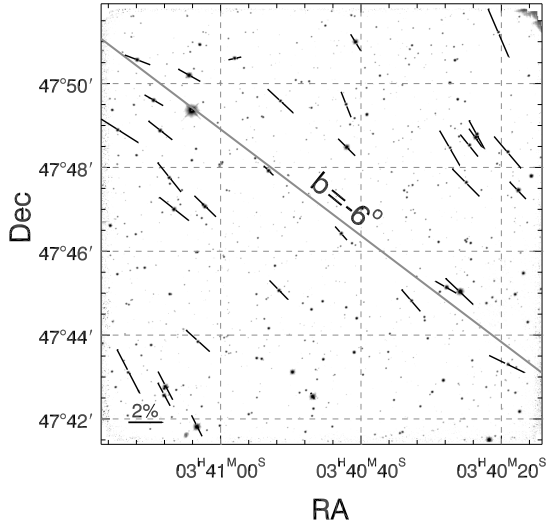}
	\caption{\label{sample-figure}Mimir coadded H-band intensity image of the $10\arcmin \times 10\arcmin$ field toward $b = -6\degr$ with overlaid significant polarization vectors. The length of each vector represents the degree of polarization relative to the 2\% scale in the lower left corner, and the vector orientation shows the equatorial polarization position angle. The diagonal light gray line shows a line of constant Galactic latitude at $b=-6\degr$.}
\end{figure}

Figure \ref{sample-figure} shows the coadded H-band intensity image obtained toward $b = -6\degr$, with measured polarizations shown as vectors. For detected polarizations, the length of the vector represents the degree of starlight polarization and its orientation represents the equatorial position angle of its electric field vector. A line of constant Galactic latitude is shown for reference.

The $\ell=150\degr$ region of sky was chosen for two reasons. First, the outer Galaxy provides a relatively quiescent region (compared to the inner Galaxy) where the large-scale magnetic field can be observed. The inner Galaxy hosts much of the Galaxy's star formation and exhibits extensive overlapping supernova remnants, which can distort the magnetic field on parsec to kiloparsec scales. Second, the $\ell=150\degr$ fields overlap completed SDSS SEGUE \citep{Y09} imaging and spectroscopy fields. Subsequent analysis of the photometry and spectroscopy obtained by SEGUE may be used to inform distance estimates for the stars observed here to enable decomposing the magnetic field projections with distance, though this is beyond the scope of this current work.

\begin{figure}
	\centering
		\plotone{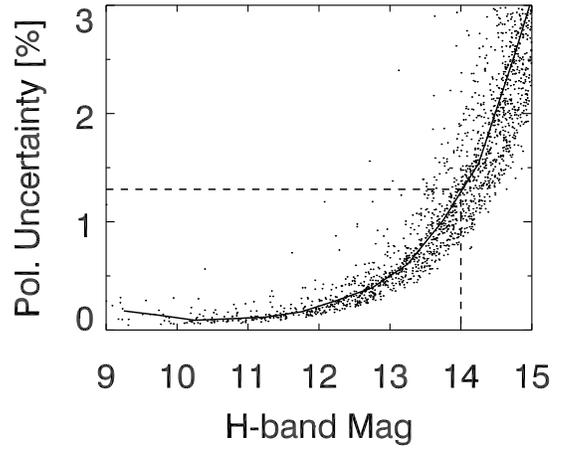}
	\caption{\label{sigma-h}Distribution of polarization uncertainty as a function of H-band magnitude. The solid line traces the median polarization uncertainty with apparent magnitude. The median polarimetric uncertainty at $H=14$ is shown by the dashed lines.}
\end{figure}

A total of 10,962 stars were detected in the combined photometric images, of which 332 had significant ($P/\sigma_P>$ 3) polarization detections, 784 had ``marginal'' ($1.25< P/\sigma_P <3$) detections, and 194 had significant upper limits ($P/\sigma_P<$ 1.25 and $\sigma_P<1\%$). All of these 1,310 polarization `targets' are listed in Table \ref{example_obs} by Galactic coordinates, degrees of polarization, GPAs, and 2MASS H- and K-band magnitudes. The lowest, median, and highest significant detected polarizations are $0.19\%$, $1.51\%$, and $12.13\%$, respectively. All polarization target coordinates were passed through SIMBAD to search for previous spectral classifications, but only eight out of the 1,310 polarization targets had entries. Of these matches, three stars had spectral classifications (A2V, A3V, and M3V) and one was identified as the X-ray source 1RXS J042738.8+555831 \citep{V99}, while the other four were only identified as stars. The SIMBAD search result suggests that this is a quiescent direction in the Galaxy suitable for probing the large-scale Galactic magnetic field.

\begin{deluxetable*}{cccccccccc}[t]
	\tablecaption{Starlight Polarization Observations \label{example_obs}}
	\tablecolumns{10}
	\tablehead{\colhead{} &
						 \colhead{} &
						 \colhead{} &
						 \colhead{} &
						 \colhead{} &
						 \colhead{} &
						 \multicolumn{4}{c}{2MASS} \\
						 \cline{7-10}
						 \colhead{$\ell$} &
						 \colhead{$b$} &
						 \colhead{P} &
						 \colhead{$\sigma_P$} &
						 \colhead{GPA} &
						 \colhead{$\sigma_{GPA}$} &
						 \colhead{H} &
						 \colhead{$\sigma_H$} &
						 \colhead{K} &
						 \colhead{$\sigma_K$} \\
						 \colhead{[deg]} &
						 \colhead{[deg]} &
						 \colhead{[\%]} &
						 \colhead{\%} &
						 \colhead{[deg]} &
						 \colhead{[deg]} &
						 \colhead{[mag]} &
						 \colhead{[mag]} &
						 \colhead{[mag]} &
						 \colhead{[mag]} \\					 
						 \colhead{(1)} &
						 \colhead{(2)} &
						 \colhead{(3)} &
						 \colhead{(4)} &
						 \colhead{(5)} &
						 \colhead{(6)} &
						 \colhead{(7)} &
						 \colhead{(8)} &
						 \colhead{(9)} &
						 \colhead{(10)} \\
						 }
	\startdata
		150.00762 & -75.07904 &   6.22 &   3.21 & 140.53 &  14.77 & 15.360 &  0.102 & 15.121 &  0.136 \\
		149.76181 & -75.05685 &   2.65 &   1.26 & 112.51 &  13.58 & 13.615 &  0.037 & 13.476 &  0.043 \\
		150.10707 & -75.02797 &   0.00 &   0.60 &   0.00 & 180.00 & 13.489 &  0.029 & 13.462 &  0.041 \\
		150.08486 & -74.97565 &   0.63 &   0.62 &  20.77 &  28.06 & 13.087 &  0.026 & 13.057 &  0.032 \\
		149.86210 & -74.96854 &   0.00 &   0.46 &   0.00 & 180.00 & 11.966 &  0.025 & 11.755 &  0.026 \\
		149.90275 & -74.95584 &   4.46 &   2.30 & 118.54 &  14.81 & 15.002 &  0.071 & 14.678 &  0.103 \\
		150.16407 & -74.95089 &   6.51 &   4.18 & 175.15 &  18.41 & 15.318 &  0.106 & 15.436 &  0.169 \\
		150.08958 & -74.93429 &   6.15 &   3.75 & 116.61 &  17.47 & 15.494 &  0.113 & 15.205 &  0.146 \\
		150.10019 & -74.93291 &   0.00 &   0.69 &   0.00 & 180.00 & 13.026 &  0.030 & 12.994 &  0.035 \\
		149.96803 & -74.92756 &   7.92 &   4.36 & 107.68 &  15.78 & 15.593 &  0.113 & 15.430 &  0.187 \\
		150.24836 & -65.02715 &   2.28 &   0.92 &  46.20 &  11.58 &  0.000 & 99.990 &  0.000 & 99.990 \\
	\enddata
	\tablecomments{Table \ref{example_obs} is published in its entirety in the electronic edition of the Astrophysical Journal. A portion is shown here
		for guidance regarding its form and content. Stars with missing 2MASS photometry are listed as having apparent magnitudes of 0.00 mag and uncertainties of 99.99 mag}
\end{deluxetable*}

To aid in determining the completeness of these observations, the distribution of H-band magnitudes for all detected stars was examined. From these distributions, the photometric observations appear complete to $H\approx 16.5$ mag and the polarimetric observations appear complete to $H\approx 13$ mag. The integration times were set so that the median polarization uncertainty, $\sigma_P$, would be less than 1\% for stars brighter than $H=14$. As shown in Figure~\ref{sigma-h}, this goal was not quite achieved, with a median polarization uncertainty of 1.3\% at $H=14$. The shape of the curve in Fig. \ref{sigma-h} indicates that polarimetric uncertainties are limited by photon noise.

\begin{figure}[b]
	\centering
		\plotone{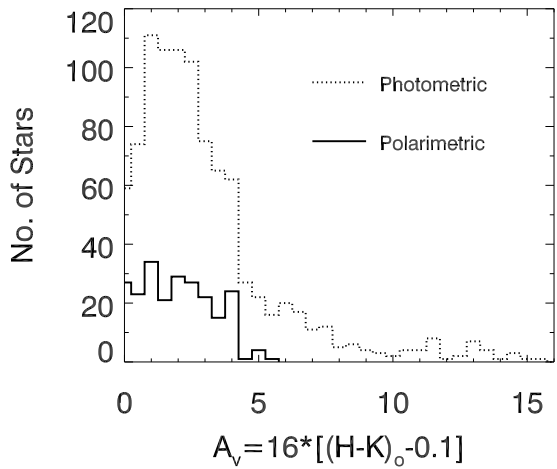}
	\caption{\label{extinction}Distributions of the estimated extinctions towards photometric (dotted line) and significantly polarized (solid line) stars, as described in the text.}
\end{figure}

The polarization efficiency ($P/A_V$) towards each star carries information about the polarization environment and polarization mechanism \citep{G92,L97,W08}. Because these observations span from the Galactic midplane to 75 degrees above the Galactic plane, a change in $P/A_V$ could indicate a change in the polarization or grain alignment mechanism. The extinction towards each star can be estimated from the observed (H-K) 2MASS colors and the intrinsic colors from \citet{BB88}. Assuming all stars have an intrinsic (H-K) color of $0.1\pm 0.1$ mag, E(H-K) values can be converted to extinctions at V-band, $A_V$, as:
\begin{equation}
	\label{av}
	A_V=r'E_{H-K},
\end{equation}
assuming $r'= 16$ for the diffuse ISM \citep{W96}. Considering only stars with measured 2MASS (H-K) colors, 2MASS photometric $SNR > 3$ for H and K, and polarimetric $SNR > 3$, the resulting photometric and polarimetric extinction distributions are shown in Figure \ref{extinction}. The median extinction of the photometric and polarimetric stars are $A_V=2.6$ and $1.8$, respectively. In Fig. \ref{extinction}, the photometric and polarimetric distributions appear to have the same overall shape, indicating that they are drawn from the same parent distribution and that the polarimetric selection criteria does not introduce bias to that sample. The observed distribution of polarization with extinction was examined for trends in the polarization efficiency ($P/A_V$) with Galactic latitude, but none were seen. This implies no detectable change in the polarization mechanism with Galactic latitude.

\begin{figure}
	\centering
		\plotone{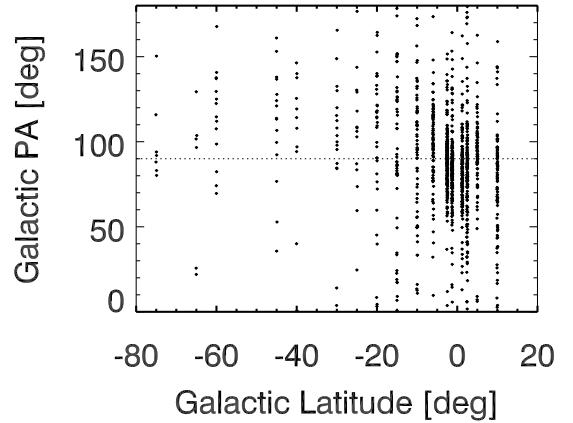}
	\caption{\label{raw_pa_vs_b}Measured Galactic position angle (GPA) as a function of Galactic latitude for stars showing significant and marginal starlight polarizations. The dotted line at GPA = $90\degr$ represents perfect alignment of the magnetic field parallel to the Galactic plane.}
\end{figure}

\begin{figure}[ht]
	\centering
		\plotone{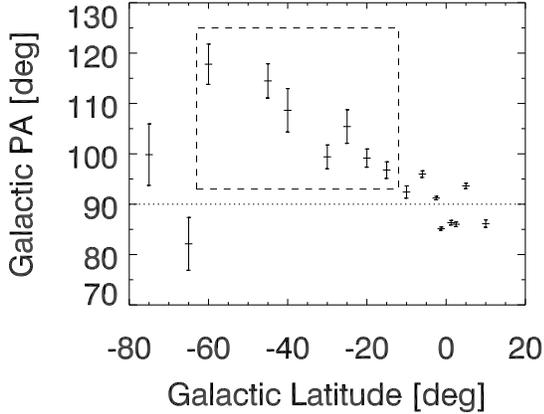}
	\caption{\label{pa_vs_b}Weighted mean GPA for each observed field, versus Galactic latitude. The error bars represent the weighted uncertainties of the means for each field, and the dotted line represents perfect alignment parallel to the Galactic plane. The dashed box identifies the Galactic latitude range that is most diagnostic for constraining the Galactic magnetic pitch angle (see text in \S4.2).}
\end{figure}

In Figure \ref{raw_pa_vs_b}, the observed GPA of each significantly and marginally detected stellar polarization is shown, as a function of Galactic latitude. The typical GPA uncertainty is $5\degr$ for the significant detections and $21\degr$ for the marginal detections. Alignment with the Galactic plane would occur at $GPA = 90\degr$, as shown by the dotted line. Fairly good alignment is seen near the Galactic midplane ($b\;\sim\;0\degr$), in agreement with the findings from optical polarimetry \citep{H2000}. To quantify the GPA for each field and how it varies with Galactic latitude, the GPA weighted means and uncertainties were calculated for each field and are shown in Figure \ref{pa_vs_b}. The error bars in the figure represent the weighted $\pm1\sigma$ uncertainties in the mean GPAs. 

\begin{figure}
	\centering
		\plotone{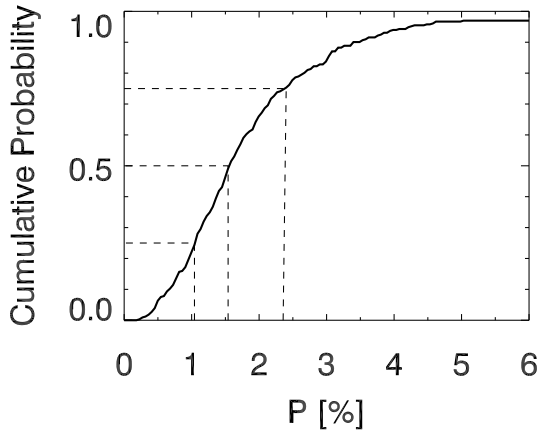}
	\caption{\label{obs_cdf}The observed starlight polarization percentage CDF for all 386 high-significance ($P/\sigma_P > 3$) stars, drawn from all latitude bins. The dashed lines show the locations of the 1st, 2nd, and 3rd quartiles of the significantly detected polarization values.}
\end{figure}

tion CDF for stars exhibiting significant polarization ($P/ \sigma_P > 3$) for all latitude bins is shown in Figure \ref{obs_cdf}. The 1st, 2nd, and 3rd quartiles occur at 1.04\%, 1.54\%, and 2.38\%, respectively, and are shown by the dashed lines in Fig. \ref{obs_cdf}. The smallest significantly detected polarization was 0.19\%, which is comparable to the polarization sensitivity limit of these observations. The polarization lower limit arises from both photon noise and calibration uncertainties, as discussed in \citet{C12b}.

\section{Comparison with Simulated Observations}

Comparisons of these observations with predictions of the large-scale structure of the Galactic magnetic field can constrain the magnetic geometry. \citet{P11} simulated all-sky NIR starlight polarization observations for several magnetic field geometries. These included: three A0 (disk antisymmetric), and three S0 (disk symmetric) magnetic field geometries \citep{FS2000}; three DEHO geometries \citep{M10}; and three analytic axisymmetric magnetic fields with different pitch angles ($\alpha = 0\degr,\, -11.5\degr,$ and $-24\degr$). In a trailing spiral galaxy, since the magnetic field is tied to the gas and not to the spiral density pattern, differental rotation causes the $B_r$ and $B_\phi$ components to have opposite signs \citep{K89}. The pitch angle is defined by $p=tan^{-1}(B_r/B_{\phi}$), and therefore the pitch angle is negative \citep{B96}. S0 and A0 magnetic fields are two common axisymmetric predictions for the large-scale structure of the Galactic magnetic field. DEHO magnetic fields, predicted by \citet{M10} and possibly observed by \citet{Sun08}, invoke a Galactic wind that connects independent halo and disk dynamos. Analytic axisymmetric magnetic fields (consisting of only $B_\phi$ and $B_r$ components) were used by \citet{P11} to predict the effects of a spiral-based Galactic magnetic pitch angle on the observed polarization distribution, and do not represent a magnetic field geometry with a known physical basis. All of these magnetic field models were combined with two different empirical dust distributions \citep{SMB96,DS01} yielding a total of 24 sets of predictions, as summarized in the first three columns of Table \ref{model_table}. In addition to an all-sky map for each prediction, an $\ell=150\degr$ GPA with $b$ plot was shown, all-sky polarization CDFs were plotted, and the 1st, 2nd, and 3rd all-sky CDF quartile values were tabulated for synthetic stellar samples complete to $H=14$.

\citet{P11} suggested two tests were most diagnostic for comparing the predictions with observations: (1) comparing the CDFs of the observed and predicted starlight polarizations, and (2) comparing the change in GPA with Galactic latitude between the observations and predictions. Each test is applied and discussed below.

\begin{deluxetable*}{ccccccccc}
	\tablecaption{Model Parameters \label{model_table}}
	\tablewidth{0pt}
	\tablecolumns{6}
	\tablehead{Model & Magnetic & Dust & K-S Test & \multicolumn{4}{c}{Fitting Constants\tablenotemark{d}} & $\overline{\Delta GPA}$ \\
						 Number & Field & Model\tablenotemark{a} & Probability & A & B & C[$\times 10^{-4}$] & D[$\times 10^{-5}$] & [deg] \\
						 (1) & (2) & (3) & (4) & (5) & (6) & (7) & (8) & (9) \\
						 }
	\startdata
			1 & S0 reference run\tablenotemark{b} & DS2001 & 0.195 & 90.1 & -0.72 & -1.53 & 4.00 & $7.14\pm0.94$  \\
			2 & S0 reference run\tablenotemark{b} & SMB96  & 0.079 &  90.1 & -0.71 & -0.98 & 3.95& $6.99\pm0.94$  \\
			3 & S0 reference with alpha quenching\tablenotemark{b} & 0.252 & DS2001 & 90.0 & -0.58 & 2.19 & 3.35 & $3.69\pm0.94$  \\
			4 & S0 reference with alpha quenching\tablenotemark{b}& 0.094 & SMB96 & 90.0 & -0.58 & 2.67 & 3.36 & $3.60\pm0.94$ \\
			5 & S0 with vacuum BC\tablenotemark{b} & DS2001 & 0.270 & 90.0 & -0.57 & 1.60 & 3.22 & $4.54\pm0.94$ \\
			6 & S0 with vacuum BC\tablenotemark{b} & SMB96 & 0.105 & 90.0 & -0.57 & 2.08 & 3.24 & $4.46\pm0.94$ \\
			\hline
			7 & A0 reference run\tablenotemark{b} & DS2001 & 0.241 & -0.20 & 0.043 & 43.4 & 9.76 & $78.12\pm1.03$ \\
			8 & A0 reference run\tablenotemark{b} & SMB96 & 0.055 & -0.03 & 0.060 & 13.0 & 7.63 & $77.04\pm1.50$ \\
			9 & A0 reference with alpha quenching\tablenotemark{b} & DS2001 & 0.241 & -0.24 & 0.039 & 41.5 & 9.58 & $77.72\pm1.08$ \\
			10 & A0 reference with alpha quenching\tablenotemark{b} & SMB96 & 0.084 & -0.12 & 0.043 & 4.19 & 6.74 & $75.91\pm1.71$ \\
			11 & A0 with vacuum BC\tablenotemark{b} & DS2001 & 0.241 & -0.27 & 0.047 & 46.0 & 10.14 & $77.75\pm1.08$ \\
			12 & A0 with vacuum BC\tablenotemark{b} & SMB96 & 0.063 & -0.19 & 0.061 & 12.6 & 7.69 & $75.94\pm1.71$ \\
			\hline
			13 & DEHO $C_{wind} = 0, R_{\alpha,halo} = 300$\tablenotemark{c} & DS2001 & $7.42\times10^{-5}$ & 90.1 & -0.80 & -14.2 & 3.31 & $8.66\pm1.15$ \\
			14 & DEHO $C_{wind} = 0, R_{\alpha,halo} = 300$\tablenotemark{c} & SMB96 & $2.70\times10^{-5}$ & 90.1 & -0.80 & -14.7 & 3.24 & $8.68\pm1.17$ \\
			15 & DEHO $C_{wind} = 100, R_{\alpha,halo} = 300$\tablenotemark{c} & DS2001 & $8.91\times10^{-5}$ & 90.1 & -0.77 & 3.87 & 5.36 & $8.69\pm1.37$ \\
			16 & DEHO $C_{wind} = 100, R_{\alpha,halo} = 300$\tablenotemark{c} & SMB96 & $4.31\times10^{-5}$ & 90.1 & -0.78 & 3.11 & 5.35 & $8.66\pm1.39$ \\
			17 & DEHO $C_{wind} = 200, R_{\alpha,halo} = 300$\tablenotemark{c} & DS2001 & $7.25\times10^{-5}$ & 89.5 & -1.13 & -61.1 & 0.64 & $14.37\pm1.28$ \\
			18 & DEHO $C_{wind} = 200, R_{\alpha,halo} = 300$\tablenotemark{c} & SMB96 & $3.32\times10^{-5}$ & 89.5 & -1.13 & -61.8 & 0.56 & $14.34\pm1.30$ \\
			\hline
			19 & Ring, $\theta=0\degr$ & DS2001 & 0.270 & 90.0 & -0.54 & 0.71 & 2.95 & $2.56\pm0.93$ \\
			20 & Ring, $\theta=0\degr$ & SMB96 & 0.167 & 90.0 & -0.54 & 1.23 & 2.95 & $2.49\pm0.94$ \\
			21 & Ring, $\theta=11.5\degr$ & DS2001 & 0.265 & 90.0 & -0.31 & 4.86 & 1.89 & $-3.58\pm0.93$ \\
			22 & Ring, $\theta=11.5\degr$ & SMB96 & 0.222 & 90.0 & -0.30 & 5.25 & 1.89 & $-3.64\pm0.94$ \\
			23 & Ring, $\theta=24\degr$ & DS2001 & 0.270 & 90.0 & -0.080 & 6.01 & 0.86 & $-9.66\pm0.93$ \\
			24 & Ring, $\theta=24\degr$ & SMB96 & 0.156 & 90.0 & -0.076 & 6.34 & 0.86 & $-9.72\pm0.94$ \\
	\enddata
	\tablenotetext{a} {DS2001 = \citet{DS01};	SMB96 = \citet{SMB96}.}
	\tablenotetext{b} {\citet{FS2000}}
	\tablenotetext{c} {\citet{M10}}
	\tablenotetext{d} {The fit is of the form $GPA = A + B\times b + C\times b^2 + D\times b^3$.}
\end{deluxetable*}

\subsection{Cumulative Distribution Functions}

The predicted CDFs in \citet{P11} were normalized in that work by the maximum predicted polarization, because of uncertainties in the degree of polarization caused by the unknown magnetic alignment and dust polarization efficiencies. If these factors do not change significantly along or between different lines of sight, then the predicted GPA is unaffected and the normalized CDFs are related to the actual polarization CDFs by a multiplicative factor. 

\begin{figure*}
	\centering
		\includegraphics[scale=0.8]{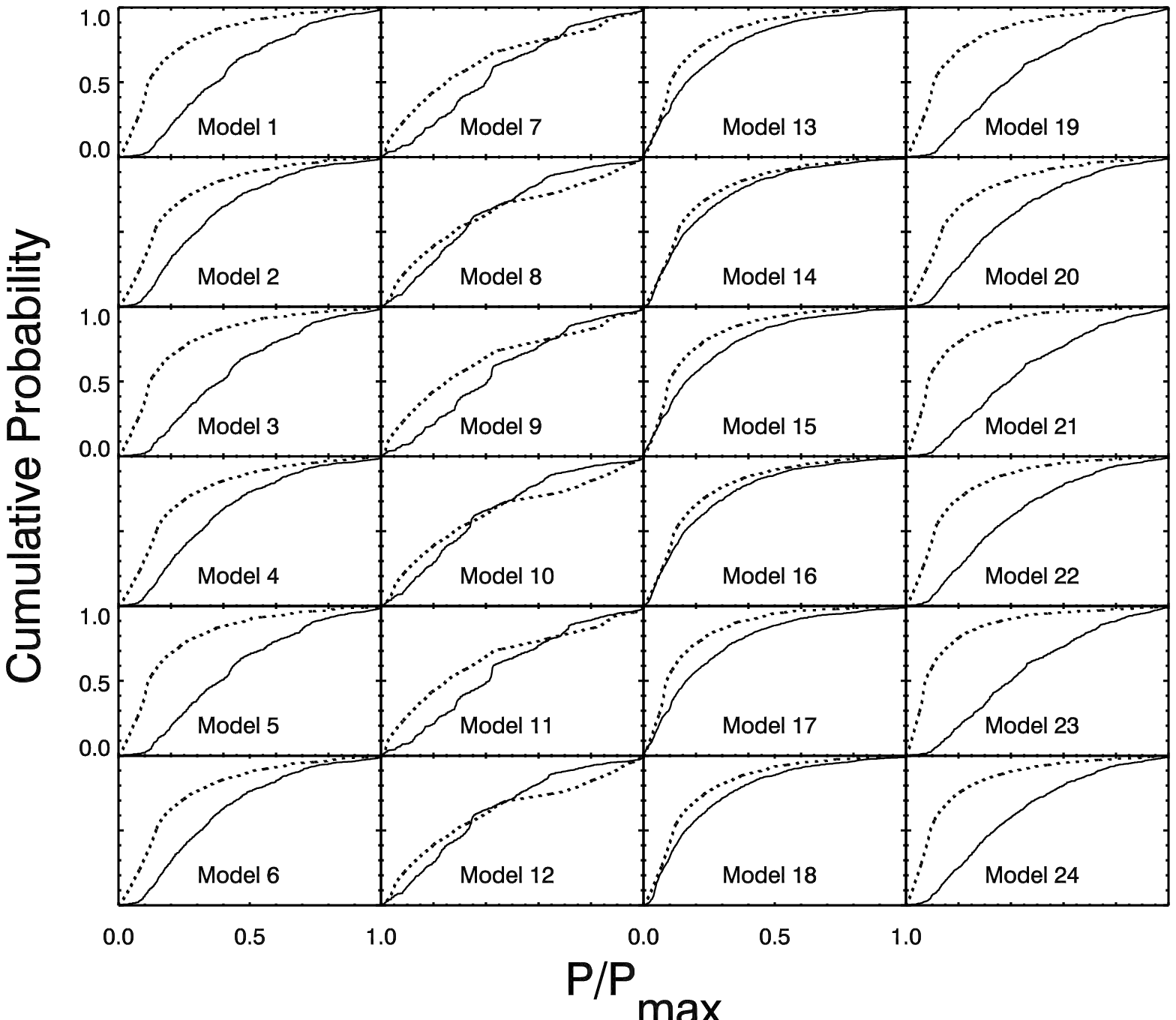}
	\caption{\label{new_cdfs}Predicted starlight polarization CDFs for the subset of directions corresponding to the observations reported here (solid lines). These curves are different from the all-sky curves presented in \citet[][dotted lines; modified as described in the text]{P11}, especially at low and high polarizations because of the limited sampling of the sky by the new observations and the latitude and longitude dependence of the polarizations.}
\end{figure*}

The \citet{P11} predictions were generated for the whole sky, while the observations presented here cover only a small region of sky. Since the predicted degree of polarization strongly depends on Galactic latitude and longitude, a subsample was drawn from the all-sky predictions which included only those directions corresponding to observations reported here. To account for the observations at $b= \pm1.25\degr$, $\pm2.5\degr$, and $-6\degr$ that have no corresponding \citet{P11} predictions, new simulations towards these directions were made, following the procedure in \citet{P11}, and included in the subsample. 

While the shapes of the variation of GPA with Galactic latitude for all 24 models presented in \citet{P11} are unchanged, the new subsample predicted polarization CDFs are, as would be expected, different from the all-sky \citet{P11} CDF predictions. In Figure \ref{new_cdfs}, the new subsample CDF predictions (solid lines), containing 332 stars, are shown along with modified all-sky CDF predictions (dotted lines) from \citet{P11}. In the original all-sky predictions, a population of distant, high polarization stars exists towards the Galactic center ($|\ell|<20\degr, b=0\degr$) that are not characteristic of the rest of the sky. The presence of these stars was identified in \citet{P11} as a discontinuity in the CDFs presented there. These very high polarization stars were removed from the all-sky CDFs shown in Fig. \ref{new_cdfs} (as dotted lines). Significant differences remain between the shapes of the modified original and subsample CDFs. The inclusion of four additional sightlines near the Galactic plane ($b=\pm1.25\degr,\,\pm2.5\degr$) to the subsamples preferentially adds higher polarization stars, because of the large dust columns expected in the plane, which increases the cumulative probability at higher polarizations. This results in a lower cumulative probability at small $P/P_{max}$ and higher probability at large $P/P_{max}$, seen in all of the CDFs as a general flattening of the subsample curves.

The multiplicative scale factors between each of the new predicted subsample CDFs and the observed CDF must also be calculated. Simple comparison of normalized predicted subsample CDFs to the observed CDF is unreliable because the maximum observed polarization value (used to normalize all other values) was based on a single star in the high end tail of the polarization probability distribution. A single extreme polarization value would affect all of the normalized starlight polarizations and change the cumulative distribution function.

For each model comparison, the null hypothesis is that the subsample-predicted and NIR-observed polarization CDFs are drawn from the \emph{same} parent distribution. If the probability that they are drawn from the same parent distribution is small, the null hypothesis can be rejected. The Kolmogorov-Smirnov (K-S) test was used to determine both the appropriate scaling factor and to perform the test of the null hypothesis. By scaling the predicted subsample CDFs so that the maximum absolute difference between the observed and predicted CDFs was \emph{minimized}, the K-S probability that the CDFs are drawn from the same parent distribution is maximized. This introduces a heavily conservative bias that overestimates the likelihood of agreement. If this \emph{overestimated probability} is still significantly small, then that particular model can be rejected. In this case, the overestimated probability is a strong, conservativly biased criteria for rejecting the null hypothesis.

For sky locations matching the observations, 866 simulated stars with H$\leq$14 mag make up the predicted subsample CDFs. The 332 observed stars with H$\leq$14 mag and significant polarizations make up the NIR-observed CDF. Many of the simulated stars, however, would actually have polarization values below the NIR observational limit and should be culled. To calculate the stars to ignore, the following procedure was used for each of the 24 subsample predictions. First, the entire predicted subsample polarization CDF was scaled so that the maximum deviation between it and the observed polarization CDF was minimized. Next, model stars with scaled polarizations below the observational polarization threshold (0.19\%) were identified and removed. Then, a new predicted polarization CDF was calculated and scaled to the observations again. The resulting (overestimated) K-S probabilities for each model-observations comparison are listed in column (4) of Table \ref{model_table}. These represent the probabilities that the best-scaled predicted CDFs were drawn from the same population as the observed CDF. The largest probability of all the models is only 0.27 and is too low to identify any obvious good match. The smallest probabilities (strongest model rejections) are approximately $10^{-5}$ for the DEHO models, though further consideration is necessary before rejection and will be discussed below.

\subsection{GPA vs. Galactic Latitude}

The second observational test proposed by \citet{P11} was based on the dependence of the polarization GPA on Galactic latitude. The observational data for this test were shown in Fig. \ref{raw_pa_vs_b}. Weighted means and uncertainties in the mean for each latitude bin were shown in Fig. \ref{pa_vs_b}. To facilitate comparison with the observations, the runs of predicted GPAs, for each model at $\ell=150\degr$, as a function of Galactic latitude \citep[see Figs. 4-7 in][]{P11} were fit by a third order polynomial:
\begin{equation}
	\label{poly}
	GPA = A + B\times b + C\times b^2 + D\times b^3
\end{equation}
where GPA is the predicted Galactic polarization position angle; $b$ is the Galactic latitude of the prediction; and A, B, C, and D are constants to be fit. The parameters from these fits are listed in columns (5)-(8) in Table \ref{model_table}. The choice of a third-order polynomial was based on an F-test. 

\begin{figure}
	\centering
		\plotone{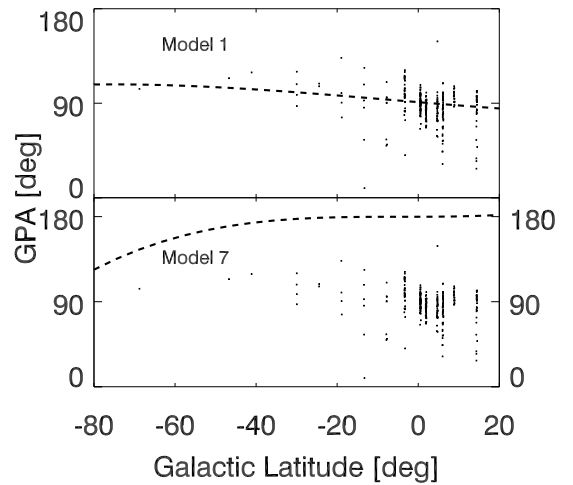}
	\caption{\label{data_and_fits}Two example model polynomial fits (dashed lines) plotted over the observed GPAs (points) as a function of Galactic latitude. For clarity, only the significantly detected (P/$\sigma_P$ $>$ 3) polarizations are plotted, and are identical in both panels. The S0-type Model 1 (top) is an example of fair agreement between the predictions and observations. The A0-type Model 7 (bottom) is an example of poor agreement, with an approximately $90\degr$ difference between the predictions and observations.}
\end{figure}

\begin{figure*}
	\centering
		\plotone{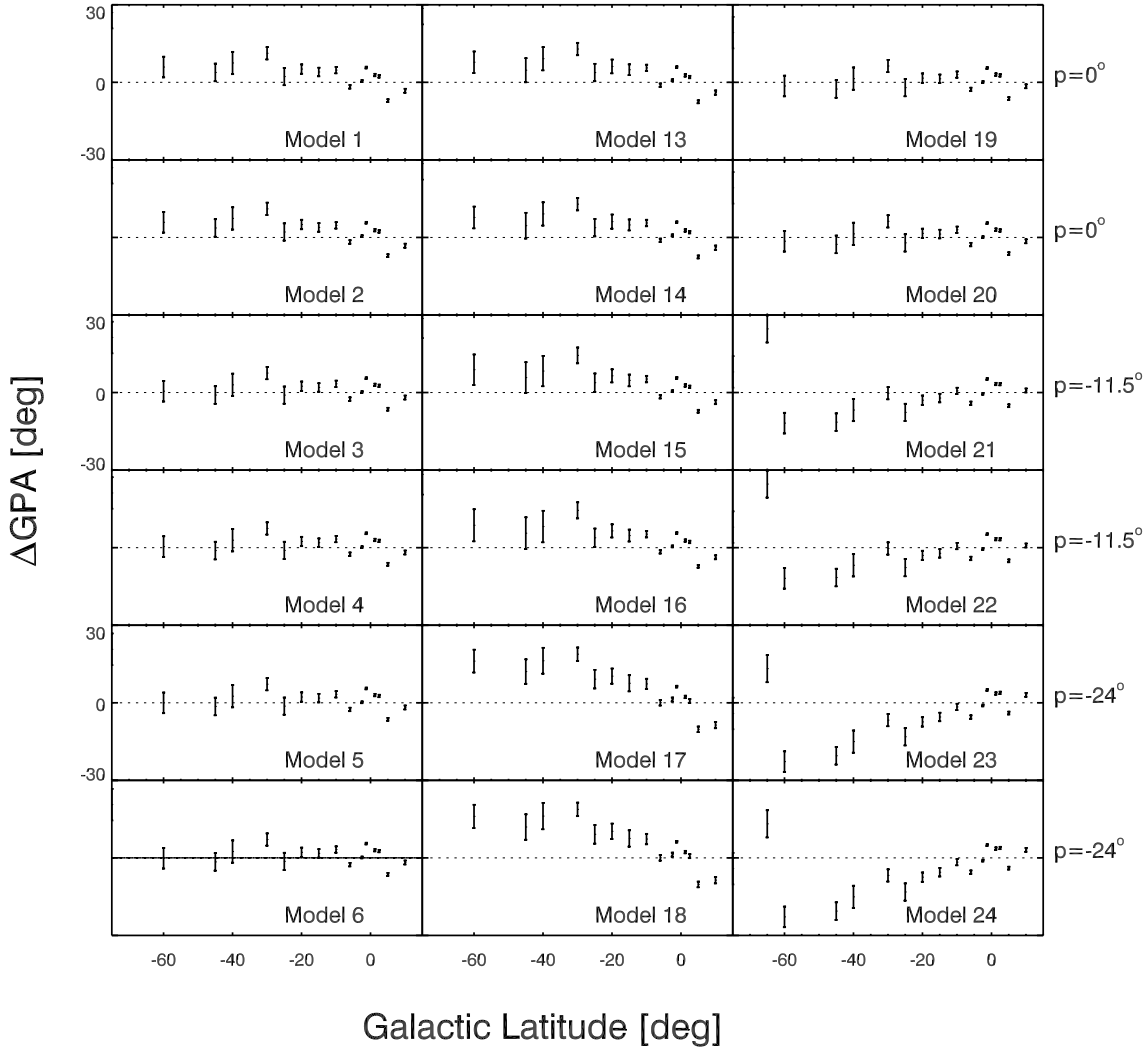}
	\caption{\label{delta_pa}The differences between the observed mean GPAs and predicted subsample mean GPAs in each latitude bin for Models 1-6 and 13-24. Models 7-12 are not shown because the differences for all latitude bins are greater than $60\degr$. The error bars show the $\pm1\sigma$ uncertainties for each difference. The effect of magnetic pitch angle on $\Delta GPA$ for $|b|>10\degr$ is seen in Models 19-24  where the pitch angle (listed to the right of each plot) is varied.}
\end{figure*}

The observed GPAs and two example model polynomial fits are plotted in Figure \ref{data_and_fits}. Model 1 (top panel) is an example of a fair match between the predictions and observations and Model 7 (bottom panel) is an example of a poor match. For each latitude bin having observations, the difference between the predicted mean GPA and observed mean GPA was calculated. These are shown in Figure \ref{delta_pa} for all models. The weighted observational mean GPA uncertainties and the predicted subsample GPA dispersions are combined in quadrature to estimate the uncertainties for these differences. In Fig. \ref{delta_pa}, Models 1-6 and 13-24 show fair agreement between predictions and observations. Models 7-12 (A0 magnetic field geometries), however, do not agree with the observations, with most latitude bins having discrepancies of $\Delta GPA > 80\degr$, essentially orthogonal to the predicted GPAs, and therefore are not shown.

To reject specific models, it is necessary to quantify these differences. For each model, a weighted average of the differences between the model's predicted GPA and observed GPA in each latitude bin, as shown in Fig. \ref{delta_pa}, was calculated and the weighted uncertainty in this average difference was obtained by propagation. These values and uncertainties are listed in column (9) in Table \ref{model_table}. While all S0, A0, and DEHO models can be rejected at the $3\sigma$ level, the A0 models are particularly discrepant ($>$40$\sigma$) since they predict polarizations essentially perpendicular to the Galactic plane.

\section{Discussion}

\subsection{Large-Scale Magnetic Field Symmetry}

Using the two tests proposed by \citet{P11} and the observations reported here, all of the magnetic field models taken from the literature can be formally rejected as deviating by more than 3$\sigma$ from the observations. The K-S analysis of CDF values was inconclusive about rejection of the predicted polarization values. None of the probabilities was high enough to suggest one set of models was the best choice, and (except for the DEHO models) the probabilities were not small enough to definitively reject classes of models outright. The differences between the predicted and observed polarization GPAs as a function of Galactic latitude are significant enough to reject all of the magnetic field models. However, the full interpretation of these results requires a more careful analysis. 

The initial results should not be completely surprising, since the actual Galactic magnetic field is unlikely to be as simply represented as the simulated magnetic field models. The dynamo-driven magnetic field models have a resolution of $\sim$200 pc \citep{M10} or $\sim$400 pc \citep{FS2000}, so any turbulence or magnetic disturbances on smaller scales could not be included. The analytic axisymmetric models, with effectively infinite resolution, also lack turbulence. If turbulence creates random magnetic field components, then along a line of sight the starlight is expected to suffer some \textit{depolarization} as it passes through multiple randomly oriented cells \citep{J89}. \citet{OS93} calculated a coherent magnetic cell size of 10 to 100 pc using Galactic pulsar rotation and dispersion measures, implying that starlight from distant stars will generally pass through many cells. \citet{P11} also discussed uncertainties in the predicted degree of polarization arising from changes in the polarization efficiency, though they are less likely to affect the GPAs. This, combined with uncertainties from the calculated scale factors, creates uncertainty in the absolute probability returned by the K-S test, though the relative probabilities may still provide insight into ranked likelihoods.

Considering the relative K-S probabilities in Table \ref{model_table}, the S0, A0, and analytic axisymmetric models using the \citet{DS01} dust distributions have the highest probability ($>$ 0.20) of being drawn from the same parent distribution as the observations. The DEHO models all produce polarization CDFs that are poor matches to the observed polarization CDF. The choice of dust model causes a significant difference in the probabilities for all models. The S0, A0, and analytic axisymmetric models using the \citet{DS01} dust distribution have larger (i.e., more similar to the observations) probabilities than the models using the \citet{SMB96} dust distribution. This is not surprising given that the \citet{DS01} dust model accounts for more details (e.g., spiral arms, local arm, Galactic warp) than the axisymmetric \citet{SMB96} dust model. The DEHO models, as shown by their K-S probabilities, are unable to reproduce the observed polarizations.

The amount of polarization increases as starlight passes through the regular (uniform) magnetic field, but depolarizes as it passes through cells of randomly oriented magnetic field. The interplay between regular and random magnetic fields is strongly dependent on the nature of turbulence, and beyond the scope of this paper. Because of the lack of depolarization in the simulations, all simulated stars will show larger polarizations than would actual stars. The exact effect on the simulated degrees of polarization requires a detailed study of the ratio of power in the random and regular magnetic fields in the diffuse ISM, analogous to the studies that have been conducted in higher density molecular clouds \citep{FG08,Hi09,H09}. The changes to the degree of polarization caused by depolarization may be large enough to change the predicted polarization CDF and account for the significant differences measured by the K-S test.

Given the uncertainties associated with the predicted degree of starlight polarization, the distribution of GPA vs. Galactic latitude may serve as a \emph{better} test for constraining the large-scale structure of the Galactic magnetic field. 

The A0 (disk-odd) Galactic magnetic fields, models 7-12, show huge differences between the observed and predicted GPAs towards $\ell=150\degr$ and are therefore not shown in Fig. \ref{delta_pa}. In typical A0 magnetic field models, the poloidal magnetic field has a dipolar configuration and the toroidal magnetic field consists of antisymmetric tori of magnetic flux above and below the Galactic disk with opposite field directions. To sustain this structure, there must be a toroidal null point approximately in the Galactic plane where the toroidal magnetic field goes to zero. At this point, only the poloidal magnetic field is expressed, which is orientated roughly in the Galactic North-South direction, orthogonal to the Galactic plane as seen in the simulations. Also, most of the toroidal magnetic flux in the specific A0 magnetic field geometries used for these simulations \citep[e.g., Fig. 10 in][for Model 7]{FS2000} is contained within the solar circle and would not be probed when looking toward the outer Galaxy. Based solely on the GPA discrepancies, these A0 models can be rejected.

The rejection of A0 magnetic field geometries is at odds with previous work utilizing rotation measure studies \citep[e.g.,][]{AM88,H97,H03,Sun08}, which probe the Galactic magnetic field over kiloparsec length lines of sight. Evidence for an antisymmetric magnetic field is seen in antisymmetric rotation measures above and below the Galactic plane towards the Galactic center. This implies that the dominant toroidal magnetic field direction is reversed above and below the Galactic plane. \citet{W2010}, however, has recently shown that this observed asymmetry may be dominated by a nearby (100 pc), northern-sky, HI bubble and that the observed asymmetry may not reflect the actual large-scale Galactic magnetic field. By using a different method (starlight polarimetry instead of Faraday rotation) away from known magnetic disturbances, the rejection of the A0 magnetic field is robust.

The analytic axisymmetric models (Models 19-24) show the best overall matches, both in degree of polarization and GPA. The difference between the observed and predicted GPAs suggests that $-24\degr$ is too large to match the observations. Based on the quality of these matches, a magnetic pitch angle between $0\degr$ and $-11.5\degr$ provides the best fit to the data.

\subsection{Magnetic Pitch Angle}

\begin{figure}
	\centering
		\plotone{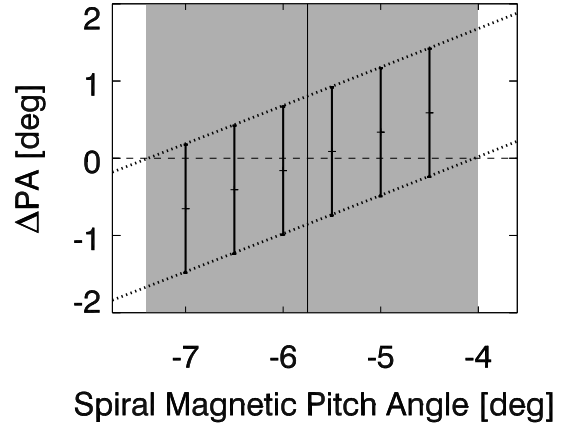}
	\caption{\label{pitch_angles}The average difference between the observed and predicted GPA distributions at $\ell=150\degr$, $-60\degr \leq b \leq -15\degr$ as a function of magnetic field spiral pitch angle for the analytic axisymmetric model. The horizontal dashed line shows where perfect agreement between the observations and predictions occurs. The error bars represent the $\pm1\sigma$ uncertainties on the mean differences. Only models with error bars enclosing $\Delta PA=0\degr$ are shown. The diagonal solid lines trace the $\pm1\sigma$ errors of the tested models. The vertical solid line is the best fit magnetic pitch angle and the grey region shows the estimated uncertainty on that value, as described in the text.}
\end{figure}

To place tighter constraints on the magnetic pitch angle, additional analytic axisymmetric model polarization subsample predictions were generated for pitch angles of $\theta=0\degr$ to $-24\degr$ in steps of $0.5\degr$. The average discrepancies between the observed and predicted GPAs were calculated for the data in the zone $-60\degr \leq b \leq -15\degr$, shown by the dashed box in Fig. \ref{pa_vs_b} (all latitudes were used in the earlier analysis). Latitude bins close to the Galactic plane were not used for two non-independent reasons: the average predicted GPAs for $|b| \leq 10\degr$ are not significantly affected by the changing pitch angle (as seen in the rightmost column of Fig. \ref{delta_pa}), and the uncertainty on the GPA differences near the midplane are much smaller than those at larger latitudes. For a properly weighted mean, the change in uncertainty with latitude will cause the non-diagnostic, low latitude bins to dominate the average while the diagnostic, large latitude bins carry little weight. Also, the observations at $b=-65\degr$ and $-75\degr$ were not used because their GPAs were not consistent with any of the models and include only one significant starlight polarization measurement (see Table \ref{obs-table}).

The average difference between the predicted and observed GPAs, and $\pm1\sigma$ uncertainties, as a function of magnetic pitch angle, are shown in Figure \ref{pitch_angles}. In this Figure, the six models whose error bars enclose $\Delta GPA=0\degr$ (horizontal dotted line) are shown. The models not shown follow the same trend away from $\Delta GPA=0\degr$ with larger and smaller model magnetic pitch angles. The diagonal solid lines trace the $\pm 1\sigma$ error bars for all of the models. The best fit magnetic pitch angle is between $\theta=-5.5\degr$ and $-6.0\degr$, shown by the vertical solid line in Fig. \ref{pitch_angles}. To estimate the uncertainty on this magnetic pitch angle, the upper and lower 1$\sigma$ uncertainty limits for all analytic axisymmetric models were each fit with second order polynomials (shown as dotted lines in Fig. \ref{pitch_angles}). The pitch angles enclosed by these fits are shown by the grey region in Fig. \ref{pitch_angles}. Based on these new observations and predictions, the Galactic magnetic pitch angle is $\theta=-6\pm2\degr$ towards $\ell=150\degr$.

Previous estimates of the Galactic magnetic pitch angle have returned similar results: $-6\degr$ \citep{Vallee88}, $-8.2\degr\pm0.5\degr$ \citep{HQ94}, $-7.2\degr\pm4.1\degr$ \citep{H96}, and $-8\degr\pm1\degr$ \citep{B07}. Unlike these previous estimates, which typically use data from large areas of the sky, this result was obtained from one Galactic longitude. Additional measurements in the outer Galaxy may show pitch angle variations with Galactic longitude that are missed when averaging over large areas.

\section{Conclusions}

New NIR starlight polarimetry observations were obtained toward the outer Galaxy that trace the orientation of the Galactic magnetic field. These observations, combined with predictions from \citet{P11} and new simulations presented here, constrain important aspects of the geometry of the large-scale Galactic magnetic field. The Kolmogorov-Smirnoff (K-S) test probabilities of the cumulative distribution functions (CDFs) of the polarization percentages and the Galactic position angle (GPA) differences can be used to formally (3$\sigma$) reject all of the models tested here, though relative rankings may also be useful. Key results from this study include:
\begin{enumerate}

\item The A0 model predictions are strongly inconsistent with the observed distribution of NIR starlight Galactic position angles (GPAs) and are thereby rejected. While here limited to a set of sample A0 models from \citet{FS2000}, the ability of any A0-type magnetic field model to reproduce the observed GPA distribution in the outer Galaxy is doubtful.

\item The GPA distributions of the S0, DEHO, and analytic axisymmetric models typically differ from the NIR starlight observations by only a few degrees. Since the model magnetic fields do not include any turbulence or systematic motions, such differences are to be expected. The general agreement shows that disk-even (S0, DEHO, and analytic axisymmetric) magnetic fields better represent the Galactic magnetic field than A0 models.

\item Based on the CDFs of observed and predicted degrees of starlight polarization and their K-S probabilities (see Table \ref{model_table}), the Disk-Even, Halo-Odd (DEHO) magnetic field models \citep{M10} are especially discrepant. This result, and all of the K-S probabilities, may be affected by uncertainties in predicting the degree of starlight polarization. 

\item Even if the predicted DEHO polarization CDF was not discrepant, the observed distribution of GPAs would still be inconsistent, though at a similar level as for the S0 and analytic axisymmetric models.

\item New simulations presented here, with different spiral-type magnetic pitch angles, when matched to the new observations, constrain the Galactic magnetic pitch angle to $-6\degr\pm2\degr$. This estimate is for a single Galactic longitude and agrees well with previous estimates of optical starlight polarization \citep{H96} and radio Faraday rotation \citep{Vallee88,HQ94,B07} that each average over large areas of the sky.

\end{enumerate}

This work highlights the utility of NIR polarimetry and its ability to place strong constraints on the Galactic magnetic field. Future application of NIR polarimetry to larger areas of the sky will provide additional constraints for Galactic dynamo theory on large scales and star formation on small scales.

\acknowledgements

This research was conducted in part using the Mimir instrument, jointly developed at Boston University and Lowell Observatory and supported by NASA, NSF, and the W.M. Keck Foundation. This publication makes use of data products from the Two Micron All Sky Survey, which is a joint project of the University of Massachusetts and the Infrared Processing and Analysis Center/California Institute of Technology, funded by NASA and NSF. This work is partially supported by NSF grants AST 06-07500 and 09-07790 and Boston University's continuing Perkins telescope partnership with Lowell Observatory.

{\it Facilities:} \facility{Perkins}


\begin{thebibliography}{}

\bibitem [Andreasyan \& Makarov(1988)]{AM88} Andreasyan, R.~R., \& Makarov, A.~N.\ 1988, Astrophysics, 28, 247 
\bibitem [Beck et al.(1996)]{B96} Beck, R., Brandenburg, A., Moss, D., Shukurov, A., \& Sokoloff, D. 1996, \araa, 34, 155
\bibitem [Beck(2007)]{B07} Beck, R.\ 2007, EAS Publications Series, 23, 19
\bibitem [Berkhuijsen(1973)] {B73} Berkhuijsen, E.M. 1973, \aap, 24, 143
\bibitem [Bessel \& Brett(1988)]{BB88} Bessel, M., \& Brett, J. 1988, \pasp, 100, 1134
\bibitem [Brandenburg et al.(1992)]{B92} Brandenburg, A., Donner, K.~J., Moss, D., et al.\ 1992, \aap, 259, 453 
\bibitem [Brandenburg et al.(1993)]{B93} Brandenburg, A., Donner, K.~J., Moss, D., et al.\ 1993, \aap, 271, 36
\bibitem [Clemens et al.(2012a)]{C12a} Clemens, D.~P., Pinnick, A.~F., Pavel, M.~D., \& Taylor, B.~F. 2012a, Submitted to ApJS
\bibitem [Clemens et al.(2012b)]{C12b} Clemens, D.~P., Pinnick, A.~F., \& Pavel, M.~D. 2012b, Submitted to ApJS
\bibitem [Clemens et al.(2007)]{C07}	Clemens, D.~P., Sarcia, D., Grabau, A., et al.\ 2007, \pasp, 119, 1385
\bibitem [Davis \& Greenstein(1951)]{DG51} Davis, L., Jr., \& Greenstein, J. L. 1951, \apj, 114, 206
\bibitem [Drimmel \& Spergel(2001)]{DS01} Drimmel, R., \& Spergel, D. 2001, \apj, 556, 181
\bibitem [Elstner et al.(1992)]{E92} Elstner, D., Meinel, R., \& Beck, R.\ 1992, \aaps, 94, 587
\bibitem [Falceta-Gon{\c c}alves et al.(2008)]{FG08} Falceta-Gon{\c c}alves, D., Lazarian, A., \& Kowal, G.\ 2008, \apj, 679, 537 
\bibitem [Ferri\`ere \& Schmitt(2000)]{FS2000} Ferri\`ere, K., \& Schmitt, D. 2000, \aap, 358, 125
\bibitem [Fosalba(2002)]{F02} Fosalba, P., Lazarian, A., Prunet, S., \& Tauber, J. A. 2002, \apj, 564, 762
\bibitem [Gehrels(1974)]{G74} Gehrels, T., ed. 1974, IAU Colloq. 23, Planets, Stars and Nebulae Studied with Photopolarimetry (Tuscon: Univ. Arizona Press)
\bibitem [Goodman et al.(1992)]{G92} Goodman, A.~A., Jones, T.~J., Lada, E.~A., \& Myers, P.~C.\ 1992, \apj, 399, 108
\bibitem [Han(2008)]{H08} Han, J. L. 2008, 
\bibitem [Han et al.(1997)]{H97} Han, J.~L., Manchester, R.~N., Berkhuijsen, E.~M., \& Beck, R.\ 1997, \aap, 322, 98 
\bibitem [Han et al.(2003)]{H03} Han, J.~L., Manchester, R.~N., Qiao, G.~J., \& Lyne, A.~G.\ 2003, Radio Pulsars, 302, 253 
\bibitem [Han \& Qiao(1994)]{HQ94} Han, J.~L., \& Qiao, G.~J.\ 1994, \aap, 288, 759 
\bibitem [Hanasz et al.(2009)]{HWK09} Hanasz, M., W{\'o}lta{\'n}ski, D., \& Kowalik, K.\ 2009, \apjl, 706, L155 
\bibitem [Hall(1949)]{HA49} Hall, J.S. 1949, Science, 109, 166
\bibitem [Heiles(1996)]{H96} Heiles, C.\ 1996, \apj, 462, 316 
\bibitem [Heiles(2000)]{H2000}	Heiles, C. 2000, \aj, 119, 923-927
\bibitem [Hildebrand et al.(2009)]{Hi09} Hildebrand, R.~H., Kirby, L., Dotson, J.~L., Houde, M., \& Vaillancourt, J.~E.\ 2009, \apj, 696, 567 
\bibitem [Hiltner(1949)]{HI49} Hiltner, W.A. 1949, Science, 109, 65
\bibitem [Houde et al.(2009)]{H09} Houde, M., Vaillancourt, J.~E., Hildebrand, R.~H., Chitsazzadeh, S., \& Kirby, L.\ 2009, \apj, 706, 1504
\bibitem [Jones(1989)]{J89} Jones, T.~J.\ 1989, \apj, 346, 728
\bibitem [Kleeorin et al.(2002)]{K02} Kleeorin, N., Moss, D., Rogachevskii, I., \& Sokoloff, D.\ 2002, \aap, 387, 453
\bibitem [Kleeorin et al.(2003)]{K03} Kleeorin, N., Moss, D., Rogachevskii, I., \& Sokoloff, D.\ 2003, \aap, 400, 9
\bibitem [Krasheninnikova et al.(1989)]{K89} Krasheninnikova, I., Shukurov, A., Ruzmaikin, A., \& Sokolov, D.\ 1989, \aap, 213, 19 
\bibitem [Lazarian et al.(1997)]{L97} Lazarian, A., Goodman, A.~A., \& Myers, P.~C.\ 1997, \apj, 490, 273 
\bibitem [Lazarian(2007)]{L07}	Lazarian, A. 2007, \jqsrt, 106, 225
\bibitem [Manchester(1974)]{Man74} Manchester, R. N. 1974, \apj, 290, 211
\bibitem [Men et al.(2008)]{MFH08} Men, H., Ferri\`ere, K., \& Han, J. L. 2008, \aap, 486, 819
\bibitem [Moss \& Sokoloff(2008)]{M08} Moss, D., \& Sokoloff, D. 2008, \aap, 487, 197
\bibitem [Moss et al.(2010)]{M10} Moss, D., Sokoloff, D., Beck, R. \& Krause, M. 2010, \aap, 512, 61
\bibitem [Ohno \& Shibata(1993)]{OS93} Ohno, H., \& Shibata, S.\ 1993, \mnras, 262, 953
\bibitem [Pavel(2011)]{P11} Pavel, M. D., 2011, \apj, 740, 21
\bibitem [Rand \& Kulkarni(1989)]{RK89} Rand, R. J., \& Kulkarni, S. R. 1989, \apj, 343, 760
\bibitem [Serkowski et al.(1975)]{SMF75} Serkowski, K., Mathewson, D.S., \& Ford, V.L. 1975, \apj, 196, 261
\bibitem [Skrutskie et al.(2006)]{S06}	Skrutskie, M.F., Cutri, R.~M., Stiening, R., et al. 2006, \aj, 131, 1163
\bibitem [Smith(1968)]{S68} Smith, F. G. 1968, \nat, 218, 325
\bibitem [Spergel, Malhotra, \& Blitz(1996)]{SMB96} Spergel, D., Malhotra, S., \& Blitz, L. 1996, in Spiral Galaxies in the Near-IR, ed. D. Minniti and H. Rix (Garching: ESO)
\bibitem [Sun et al.(2008)]{Sun08} Sun, X. H., Reich, W., Waelkens, A., \& En$\beta$lin, T. A. 2008, \aap, 477, 573
\bibitem [Vallee(1988)]{Vallee88} Vallee, J.~P.\ 1988, \aj, 95, 750
\bibitem [Voges et al.(1999)]{V99} Voges, W., Aschenbach, B., Boller, Th., et al. 1999, \aap, 349, 389
\bibitem [Whittet(1992)]{W92} Whittet, D.C.B., Martin, P.~G., Hough, J.~H., et al. 1992, \apj, 386, 562
\bibitem [Whittet et al.(1996)]{W96} Whittet, D.C.B., Smith, R.~G., Adamson, A.~J. et al. 1996, \apj, 458, 363
\bibitem [Whittet et al.(2008)]{W08} Whittet, D.~C.~B., Hough, J.~H., Lazarian, A., \& Hoang, T.\ 2008, \apj, 674, 304 
\bibitem [Wolleben et al.(2010)]{W2010} Wolleben, M., Fletcher, A., Landecker, T. L., et al. 2010, \apj, 724, 48
\bibitem [Yanny et al.(2009)]{Y09} Yanny, B, et al. 2009, \aap, 137, 4377
\end{thebibliography}
\end{document}